\documentclass{elsart}

    \usepackage{latexsym,bm,amsmath,amssymb,amsfonts}
    \usepackage{epsfig,graphics,graphicx}
    \usepackage{makeidx}

    \renewcommand{\theequation}{\thesection.\arabic{equation}}

\long\def\comment#1{ }

\newcommand{\beq}{\begin{eqnarray}}
\newcommand{\eeq}{\end{eqnarray}}
\newcommand{\be}{\vspace{-.4cm}\begin{eqnarray}}
\newcommand{\ee}{\vspace{-.5cm}\end{eqnarray}}

\newcommand{\BQ}{\begin{equation}}
\newcommand{\EQ}{\end{equation}}
\newcommand{\BQA}{\begin{eqnarray}}
\newcommand{\EQA}{\end{eqnarray}}

\def\simge{\mathrel{%
   \rlap{\raise 0.511ex \hbox{$>$}}{\lower 0.511ex \hbox{$\sim$}}}}
\def\simle{\mathrel{
   \rlap{\raise 0.511ex \hbox{$<$}}{\lower 0.511ex \hbox{$\sim$}}}}

\begin{document}


\begin{frontmatter}

\parbox[]{16.0cm}{ \begin{center}
\title{Jet fragmentation and gauge/string duality}

\author{Yoshitaka Hatta and Toshihiro Matsuo  }

\address{  Graduate School of Pure and Applied Sciences, University
of Tsukuba, \\Tsukuba, Ibaraki 305-8571, Japan
}



\begin{abstract}
We consider an analog of $e^+e^-$ annihilation in gauge theories which have a dual string description in asymptotically $AdS_5$ space and discuss the nature of jet fragmentation. We construct the timelike anomalous dimension which governs the scale dependence of  the fragmentation function. In the limit of infinite 't Hooft coupling, the average multiplicity rises linearly with the energy and the inclusive spectrum is peaked at the kinematical boundary.
\end{abstract}
\end{center}}

\end{frontmatter}

\section{Introduction}
In this article we address the issue of jet fragmentation in ${\mathcal N}=4$ supersymmetric Yang--Mills theory (SYM) and in related theories at large 't Hooft coupling $\lambda \gg 1$.
  More precisely, since the notion of a jet may not exist at strong coupling in the usual sense, we actually discuss the fate of a timelike `photon'  created in a high energy process analogous to $e^+e^-$ annihilation.
  In QCD the corresponding problem has been traditionally a field of intense research where many ingenious theoretical predictions have been tested and verified in experiments with remarkable precision. (See, e.g., \cite{Ellis:1991qj,Kluth:2006bw}.) Our motivation to study a similar problem at strong coupling  largely comes from the recent interesting developments in understanding high energy processes in ${\mathcal N}=4$ SYM and related theories  using the AdS/CFT correspondence \cite{Maldacena:1997re}. In particular, deep inelastic scattering (DIS) has been conceived and investigated in this framework \cite{Polchinski:2002jw,Hatta:2007he,Brower:2007xg,Hatta:2007cs,BallonBayona:2007qr,Cornalba:2008qf,Cornalba:2008sp}. $e^+e^-$ annihilation is related to DIS via crossing symmetry, and therefore one naturally expects that the essential features of the strong coupling dynamics extracted from  DIS reappear in some form and leave a significant impact on the structure of jets, or even the very existence of jets. In an attempt to elucidate this point, Ref.~\cite{Hatta:2008tx} studied the time evolution of a virtual photon falling down in (Schwarzschild--)$AdS_5$ whose dual image on the gauge theory side was argued to be a jet--like branching tree. (See, also, \cite{Chesler:2008wd}.) Here we employ a field--theoretical approach developed for jet physics and make crucial use of a deep reciprocal relation between the spacelike and timelike processes in order to shed light on the small--$x$ regime of jet structure. As we shall see, the picture of fragmentation emerging from our analysis is very intuitive and serves as a nontrivial test of the scenario envisaged in \cite{Hatta:2008tx}.

   Among the many observables to characterize the final state in $e^+e^-$ annihilation, we shall focus on the average multiplicity and inclusive spectra as they are very sensitive to the details of the fragmentation process prior to the hadronization stage.   Another observable--the energy correlation function--has been recently studied in \cite{Hofman:2008ar}, with which our results are consistent.
In Section~2, we derive the timelike anomalous dimension in ${\mathcal N}=4$ SYM at strong coupling which is a key ingredient  to characterize the fragmentation process. Using this, in Section~3 we study the energy dependence of the average multiplicity and the inclusive distribution of the fragments. We provide a physical interpretation of the results and compare them with the weak coupling behavior in QCD.  Finally, we conclude in Section~4.

\section{Timelike vs. spacelike anomalous dimensions}
It has been known for quite some time that there exists a deep relation between the timelike ($T$) and spacelike ($S$)  anomalous dimensions which govern the scale dependence of fragmentation functions and parton distributions, respectively  \cite{Drell:1969jm,Gribov:1972rt}. In its modern and most elegant form, the statement is that the two types of anomalous dimensions can be derived from a single function as \cite{Dokshitzer:2005bf,Dokshitzer:2006nm,Basso:2006nk}
\begin{align}
\gamma_S(j)=f(j-\gamma_S(j))\,, \label{dok} \\
\gamma_T(j)=f(j+\gamma_T(j))\,, \label{dok2}
\end{align}
where $j$ is the Lorentz spin and, for simplicity, the beta--function is assumed to be zero.\footnote{Our normalization of anomalous dimensions is  such that the evolution equation in the Mellin space reads
\begin{align}\partial_{\ln Q^2} D_{S/T}(j,Q^2) = \gamma_{S/T}(j)D_{S/T}(j,Q^2)\,, \label{1}
\end{align}
 where
 \begin{align}
 D_{S/T}(j,Q^2)=\int_0^1 dx \, x^{j-1}D_{S/T}(x,Q^2)\,,
 \end{align}
 is the $j$--th moment of the parton distribution/fragmentation function and $x$ is the Bjorken/Feynman variable.
 }
  (\ref{dok})--(\ref{dok2}) can be easily inverted to give
\begin{align}
f(j)=\gamma_S(j+f(j))=\gamma_T(j-f(j))\,, \label{recip}
\end{align}
or equivalently,
\begin{align}
\gamma_S(j)=\gamma_T(j-2\gamma_S(j))\,,  \label{nao} \\
\gamma_T(j)=\gamma_S(j+2\gamma_T(j))\,.
\label{rec} \end{align}
The above reciprocal relation was  first recognized in the form (\ref{rec}) in the context of the QCD coherence effect near $j\approx 1$ \cite{Mueller:1983js}. Subsequently it has passed many consistency checks up to three--loops \cite{Mitov:2006ic} and been used to make predictions both at small and large $j$ in QCD and its supersymmetric extensions. In particular, $\mathcal{N}=4$ SYM, having vanishing beta--function, offers an ideal testing ground for the physics consequences of (\ref{dok}) \cite{Dokshitzer:2006nm,Basso:2006nk,Beccaria:2007bb,Beccaria:2008fi}. In the context of the AdS/CFT correspondence,
   the authors of \cite{Basso:2006nk}  discussed an application of the relation at large $j\gg \sqrt{\lambda}$  where  one can calculate $\gamma_S(j)$ by semiclassically computing the energy and the spin of a string rotating in $AdS_5$  \cite{Gubser:2002tv}. The outcome of their analysis lends  support to the idea that  the relation works even at strong coupling. (See a footnote to (\ref{ff}) below.)  In the following, we conjecture that this is indeed the case and discuss its physical consequences for jet fragmentation. Perhaps we can justify our procedure better \emph{a posteriori } by showing that the result we shall get is very physical and perfectly matches what one would expect to occur when the coupling is large.

 We shall be mostly interested in the small $j$ region, $j \sim {\mathcal O}(1)$, or equivalently, the small Feynman/Bjorken--$x$ region, $x\ll 1$.  In the spacelike case, this region is relevant to high energy, small angle scattering in gauge theory. The corresponding anomalous dimension $\gamma_S(j)$
   is known to lowest order in $1/\sqrt{\lambda}$ \cite{Kotikov:2004er,Brower:2006ea} and reads\footnote{See \cite{Cornalba:2007fs} for a discussion of higher order terms.}
\begin{align}
\gamma_S(j)=-\frac{1}{2}\left(\sqrt{2\sqrt{\lambda}(j-j_0)}-j\right)\,, \label{lip}
\end{align}
where $j_0=2-2/\sqrt{\lambda}$ is the strong coupling Pomeron intercept. (As $\lambda\to \infty$, it reduces to the graviton intercept $j_0=2$.) Note that $\gamma_S(2)=0$ as required by the energy conservation of parton distributions. Though (\ref{lip}) was originally derived for $j\approx 2$, its range of validity is arguably  much wider $\sqrt{\lambda} \gg j \ge j_0$ at least for the dominant term \cite{Kotikov:2004er,Hatta:2007he}.
Using this relation in (\ref{nao}), we find
\begin{align}
\gamma_S(j)=\gamma_T\left(\sqrt{2\sqrt{\lambda}(j-j_0)}\right)=-\frac{1}{2}\left(\sqrt{2\sqrt{\lambda}(j-j_0)}-j\right)
\,, \label{sign}
\end{align}
which can be easily solved for $\gamma_T(j)$
\begin{align}
\gamma_T(j)=-\frac{1}{2}\left(j-j_0-\frac{j^2}{2\sqrt{\lambda}}\right)\,. \label{con}
\end{align}
 (\ref{con}) automatically satisfies (\ref{rec}), and vanishes at $j=2$ as a consequence of the energy conservation of the fragmentation function.
  We see that while $\gamma_S$ is typically large $\sim \lambda^{1/4}$ away from $j=2$, $\gamma_T$ remains of the same order  as  $j$. This is not in contradiction to the usual prediction of the AdS/CFT correspondence, since $\gamma_T$ is not related to the anomalous dimension of some operator in gauge theory.
  Note that, unlike (\ref{lip}), (\ref{con}) can be safely continued to $j\le j_0$. In the spacelike case, $j$ is bounded from below $j\ge j_0$, whereas in the timelike case $\gamma_T$ is (formally) bounded from below by the minimum value $\gamma_T(\sqrt{\lambda}) \approx -\sqrt{\lambda}/4$. (Remember, however, that (\ref{con}) may not be trusted for such a large value of $j$.)
The universal function $f(j)$ can also be derived
 \begin{align}
 f(j)=j+\sqrt{\lambda}-\sqrt{4\sqrt{\lambda}j+\lambda-4\sqrt{\lambda}+4}\,, \label{ff}
 \end{align}
 though its precise form is not very illuminating, unlike the $j\gg \sqrt{\lambda}$ case where $f(j)$ exhibits a remarkable regularity when suitably expanded \cite{Basso:2006nk}.\footnote{Ref.~\cite{Basso:2006nk} demonstrated this  at strong coupling for large twist operators and kept the corresponding property for the twist--two operator implicit. Using  the result of \cite{Gubser:2002tv} for the twist--two anomalous dimension $\gamma_S(j)$, we have checked that  the function $f(j)$
  admits a large--$j$ expansion of the form
  \begin{align} f(j)=A\ln j(j+1) + B+\mathcal{O}\left(\frac{1}{j(j+1)}\right)\,, \end{align}
where, unlike $\gamma_S(j)$, terms of order $\sim 1/j$, $(\ln j)/j$ are absent to relevant order in $1/\sqrt{\lambda}$. This is a manifestation of the Gribov--Lipatov reciprocity \cite{Gribov:1972rt} which on general grounds \cite{Dokshitzer:2006nm} suggests that even at strong coupling the timelike and spacelike processes are related as described by the relations (\ref{dok})--(\ref{dok2}).}

\section{Jet fragmentation at strong coupling}
Consider an analog of $e^+e^-$ annihilation in ${\mathcal N}=4$ SYM and create a timelike photon with virtuality $Q^2$. Such a photon can be accommodated in SYM by gauging a U(1) subgroup of the ${\mathcal R}$--symmetry group. We then suppose that the theory is provided with  an infrared scale $\Lambda$. This may represent an actual deformation of the original  theory  \cite{Polchinski:2002jw}  if one is interested in applications to confining theories.
          Otherwise it is simply a parameter at which we decide to stop the evolution. At high energy such that $Q\gg\Lambda$,  the photon fragments into ${\mathcal N}=4$ SYM quanta
        (`partons')\footnote{It is one of the main points of Refs.~\cite{Hatta:2007he,Hatta:2007cs} to argue that a partonic picture is possible even at strong coupling. Indeed, structure functions computed in the AdS/CFT framework typically exhibit
saturation at low--$x$ and  admit the usual interpretation as the phase space density of partons.}
via successive branchings until the scale $\Lambda$ is reached. The early stages of the evolution is governed by the conformal dynamics of the original theory and this allows one to  make several  predictions which are unaffected by the details of the hadronization process.
  In particular, the average multiplicity of the decay particles is governed by the timelike anomalous dimension  $\gamma_T(j)$ at $j=1$ (see, e.g., \cite{Ellis:1991qj}). In a conformal theory,
\begin{align}
n(Q^2) \propto Q^{2\gamma_T(1)}\,.
\end{align}
Using (\ref{con}), we find
\begin{align}
n(Q^2)\propto Q^{1-3/2\sqrt{\lambda}}\,. \qquad (\lambda \gg 1) \label{str}
\end{align}
  Let us compare this result with that at weak coupling $\lambda \ll 1$ where the calculation of $\gamma_T(1)$ requires a careful resummation of  many angular--ordered diagrams \cite{Bassetto:1979nt,Mueller:1981ex}. Since the process is gluon dominated, the result for QCD can be readily taken over to ${\mathcal N}=4$ SYM just by turning off  the  running of the coupling
\begin{align}
n(Q^2)\propto Q^{\sqrt{\frac{\lambda}{2\pi^2}}}\,. \qquad (\lambda \ll 1)
\end{align}
  The nontrivial square--root behavior is a manifestation of the intricate quantum coherence effect in jet fragmentation which can nevertheless be interpreted  semiclassically  as a branching process  \cite{Marchesini:1983bm}. On the other hand, in the strong coupling limit the corresponding picture, as recently proposed in \cite{Hatta:2008tx} based on an analysis of the supergravity equation of motion, is surprisingly simple:
At strong coupling it is reasonable to assume that the momentum is more or less democratically divided by the decay products in each step of the $1 \to 2$ branchings since \emph{a priori} there is no reason to put emphasis on soft or collinear emissions. Then the typical energy of a parton in the $i$--th generation is
\begin{align}
 Q_i=\frac{Q}{2^i}\,. \label{spi}
\end{align}
The branching stops after $N$ steps when the energy reaches the infrared scale $\Lambda$
\begin{align}
Q_N=\frac{Q}{2^N}\sim \Lambda\,.
\end{align}
Then the total multiplicity is
\begin{align}
n(Q^2) \sim 2^N \sim \frac{Q}{\Lambda}\,, \label{frac}
\end{align}
in agreement with the $\lambda \to \infty$ limit of (\ref{str}). This suggests that at strong coupling branching is so efficient that all the fragments in the final states have the minimal energy fraction (the Feynman--$x$) proportional to their mass
 \begin{align}
x\sim \frac{\Lambda}{Q}\,. \label{ob}
\end{align}
  We recall that a closely related phenomenon occurs in the spacelike case (DIS) where it was found that, in the strict large $N_c$ limit, most of the partons have a very small energy fraction (the Bjorken--$x$)  \cite{Polchinski:2002jw,Hatta:2007he}
 \begin{align}
 x \sim \left(\frac{\Lambda}{Q}\right)^{\frac{\sqrt{\lambda}}{2}}\,.
 \end{align}

The characteristic scale (\ref{ob}) comes out naturally also from an analysis of the inclusive distribution.
 To see this, let us return to (\ref{con}) and take the supergravity limit $\lambda \to \infty$
\begin{align}
\gamma_T(j)\approx 1-\frac{j}{2}\,.
\end{align}
This means that, in the Mellin space,
\begin{align}
D_T(j,Q^2)=\int_0^1 dx\, x^{j-1} D_T(x,Q^2)=D_T(j,\Lambda^2)\left(\frac{Q^2}{\Lambda^2}\right)^{\gamma_T(j)}=
D_T(j,\Lambda^2)\left(\frac{Q}{\Lambda}\right)^{2-j}\,. \label{Q}
\end{align}
  Without detailed knowledge of $D_T(j,Q^2)$ it is not possible to explicitly invert (\ref{Q}). Still one can easily see that, at high energy $Q/\Lambda \gg 1$, the solution will be of the form
\begin{align}
D_T(x,Q^2)=\frac{Q^2}{\Lambda^2} F\left(\frac{Q}{\Lambda}x\right) \label{we}
\end{align}
where the unknown function $F(y)$ is a rapidly decaying (faster than a power law) function when $y>1$ so that the upper limit $x=1$ of the $x$--integration in (\ref{Q}) is irrelevant. Therefore, the fragmentation function, hence the inclusive distribution
\begin{align}
\frac{x}{\sigma}\frac{d\sigma}{dx}\sim xD_T(x,Q^2)\,,
\end{align}
is peaked around the kinematic boundary
\begin{align}
x\sim \frac{\Lambda}{Q}\,,  \label{low}
\end{align}
 in agreement with (\ref{ob}). Let us view the evolution (\ref{Q}) in the  $x$--space by performing an inverse
 Mellin transformation
 \begin{align}
 \frac{\partial}{\partial \ln Q^2} D_T(x,Q^2)=\int^1_x \frac{dz}{z} \,P(z) D_T\left(\frac{x}{z},Q^2\right)\,,
 \end{align}
 where the kernel $P(z)$ is a distribution localized at $z=1$\footnote{Though it is tempting to regard $P(z)$ as a `splitting function', this is not the same as the actual splitting probability in the branching picture of \cite{Hatta:2008tx} (see, (\ref{spi})). For instance, in weak coupling QCD the resummed anomalous dimension $\gamma_T$ \cite{Bassetto:1979nt,Mueller:1981ex} bears little resemblance to the lowest order splitting probability occurring in the
 branching process  \cite{Marchesini:1983bm}. }
 \begin{align} P(z)=\delta(1-z)+\frac{z}{2}\frac{\partial}{\partial z} \delta(1 -z)\,.
 \end{align}
  One sees that the evolution of $D_T(x,Q^2)$ is entirely driven by $D_T$ and its derivative at the same value of $x$, and not at larger values $x/z > x$. This is simply because  there are essentially no partons at larger $x$. All the partons have the  minimal possible energy $E\sim \Lambda$, and as $Q$ increases, they are redistributed towards a smaller $x$ region due to the trivial $Q$ dependence in the definition
  \begin{align}
  x=\frac{2E}{Q}\sim \frac{\Lambda}{Q}\,.
  \end{align}

 Again it is interesting to compare the above result with that in weak coupling QCD. The inclusive distribution
is roughly an inverse Gaussian in $\ln 1/x$ peaked around a parametrically larger value of $x$ \cite{Mueller:1981ex,Dokshitzer:1982fh}
\begin{align}
x\sim \sqrt{\frac{\Lambda}{Q}} \gg \frac{\Lambda}{Q}\,.
\end{align}
This is a striking consequence of the coherent angle--ordered radiation which depresses the spectrum around the kinematic limit (\ref{low}). Apparently the coherence is lost and large angle emissions are not suppressed at strong coupling.

For larger values of $x$, our prejudice, though we do not have a means of actually computing it, would be an exponential falloff
\begin{align}
D_T(x,Q^2) \sim \exp\left[  -\left(\frac{Q}{\Lambda}x\right)^a\right]\,, \qquad (x\gg \Lambda/Q) \label{tun}
\end{align}
 where $a$ is a positive number possibly depending on  ranges of $x$ values.\footnote{When $\lambda$ is large but finite, a new structure appears in the large--$x$ region $1/(1-x)\gg \sqrt{\lambda}$ where the evolution is governed by the so--called cusp anomalous dimension with
  $j\sim 1/(1-x)\gg \sqrt{\lambda}$. The reciprocity relation and the
result of \cite{Gubser:2002tv} combine to give
\begin{align}
\gamma_T(j) \approx \gamma_S(j)\approx -\frac{\sqrt{\lambda}}{2\pi} \ln j\,.
\end{align}
This leads to a very strong  suppression
\begin{align}
D_{S/T}(x,Q^2) \propto (1-x)^{\frac{\sqrt{\lambda}}{2\pi}\ln Q^2-1}\,, \quad (x\approx 1)
\end{align}
which completely evacuates partons in the large--$x$ region. As argued in \cite{Polchinski:2002jw}, the structure function
 in this regime is rather dominated by certain higher twist operators which we do not discuss here.}
 (\ref{tun}) is formally the same as what comes out of a supergravity calculation of the structure function in DIS at finite temperature if one replaces the Feynman--$x$  with the Bjorken--$x$ and identify $\Lambda$ with the temperature $T$. ($a=1/2$ in that case. See, (B.11) of \cite{Hatta:2007cs}.)  There the only possibility to find partons at $x>T/Q$ is via quantum tunneling which typically leads to an exponential  suppression (\ref{tun}). Note that this analogy is made possible due to a curious coincidence of the `saturation momentum' $Q_s(x)\sim 1/x$ in (\ref{low}) and Ref.~\cite{Hatta:2007cs}.

\section{Conclusion}
Recently, the angular distribution of energy and its correlations created by the static ${\mathcal R}$--current
 have been studied in detail in \cite{Hofman:2008ar}. It was found that the average energy distribution is spherical for all values of the coupling $\lambda$ in ${\mathcal N}=4$ SYM. However, on event--by--event basis the final states look very different depending on $\lambda$.  One way to see this is to study the small angle limit of the energy two--point correlation function \cite{Hofman:2008ar}
 \begin{align}
 \langle {\mathcal E}(\Omega_1){\mathcal E}(\Omega_2)\rangle \sim \frac{1}{|\theta_{12}|^{2+2\gamma_S(3)}}\,.
 \end{align}
At weak coupling, $\gamma_S(3)\sim {\mathcal O}(\lambda)\ll 1$, so the correlator is singular as $\theta_{12}\to 0$ and this   indicates the existence of collimated energy packets (jets) in the final state. The distribution is nevertheless spherical because the averaged angular dependence of jets   initiated from Weyl fermions $\sim 1+\cos^2 \theta$ is canceled by that from scalars $\sim \sin^2\theta$. On the other hand, at strong coupling $\gamma_S(3)\approx -\lambda^{1/4}/\sqrt{2}$ from (\ref{lip}), and therefore $2+2\gamma_S(3)<0$. In this case two particles arriving at nearby points in the phase space are  uncorrelated and are likely to come from different branches of the evolution tree through wide angle emissions which, as we have argued, are not suppressed at strong coupling. Combining these observations with our results on the fragmentation function, we corroborate the previous conclusion \cite{Hatta:2008tx,Hofman:2008ar,Lin:2007fa,Strassler:2008bv} that there are no jets in the final state at strong coupling. Rather, all the $Q/\Lambda$ particles have the minimal four--momentum $|p^\mu| \sim \Lambda$ and are spherically distributed in each event. A possible relationship between this picture and statistical models of particle production in $e^+e^-$ annihilation will be studied elsewhere \cite{hatta}.

\section*{Acknowledgments}
We thank Al Mueller for a helpful comment. This work is supported, in part, by
Special Coordination Funds for Promoting Science and Technology of
the Ministry of Education, Culture, Sports, Science and
Technology, the Japanese Government.

\end{document}